\documentclass{aa}  

\usepackage{graphicx}
\usepackage{multicol}
\usepackage{txfonts}
\usepackage{hyperref}
\newcommand\ed{\tilde{\delta}}

\usepackage{xcolor}

\begin{document}

\title{Hubble Expansion Signature on Simulated Halo Density Profiles:}
\subtitle{A Path to Observing the Turnaround Radius}
   \author{Giorgos Korkidis
         \inst{1}\fnmsep\inst{2}
          \and
          Vasiliki Pavlidou 
  \inst{1}\fnmsep\inst{2}
          }
    
    \authorrunning{ Korkidis  \&  Pavlidou}
    \titlerunning{Hubble expansion signature on simulated halo density profiles}

   \institute{Department of Physics and Institute for Theoretical and Computational Physics, University of Crete, GR-70013 Heraklio, Greece
   email:{\tt{gkorkidis@physics.uoc.gr;pavlidou@physics.uoc.gr}}
         \and
             Institute of Astrophysics, Foundation for Research and Technology – Hellas, Vassilika Vouton, GR-70013 Heraklio, Greece
             }

   \date{}

 
  \abstract
   {Density profiles are important tools in galaxy cluster research, offering insights into clusters dynamical states and their relationship with the broader Universe. While these profiles provide valuable information about the matter content of the Universe, their utility in understanding its dark energy component has remained limited due a lack of tools allowing us to study the transition from cluster portions that are relaxed and infalling, to those that are merging with the Hubble flow.}
   {In this work we investigate signatures of this transition in stacked density profiles of simulated cluster-sized halos at different redshifts.}
   {To highlight the Hubble flow around clusters we use their turnaround radius to normalize stacked simulated density profiles and calculate their logarithmic slope. Then, we complement our analysis by modeling the outer portions of these profiles assuming Gaussian early Universe statistics and spherical collapse without shell-crossing.}
   {We find the logarithmic slope of median cluster density profiles beyond the turnaround radius - where the Hubble flow dominates - to be Universal and well described by our model. Importantly, we find the slope of the profiles to diverge from the SCM prediction from within the turnaround radius where the actual profiles exhibit caustics which give rise to the splashback feature.}
   {We suggest utilizing this divergence from the spherical collapse model as a method to identify the turnaround radius in stacked cluster density profiles, offering a new perspective on understanding cluster dynamics and their cosmological implications.}


   \maketitle
%

\section{Introduction} \label{introduction}

Density profiles of halos are fundamental to cosmological research. Their systematic analysis, first through analytical methods \citep{GG1972, FG1984, B1985, Density_profiles_theory1} and later through N-body simulations \citep{Density_profiles_sims0, Density_profiles_sims1, Density_profiles_sims2, Density_profiles_sims3} and fitting functions \citep{Einasto, Hernquist1990, NFW, Diemer_profile}, has been crucial in advancing our understanding of the internal dynamics of galaxy clusters and their connection to the cosmic web. A consistent finding across these studies is the emergence of universal characteristics, despite the observable variability among individual structures.

This consistency is accompanied by variations that are significant for understanding dynamic transitions within clusters, notably virialization and splashback. Virialization is theorized to take place at a radius where a structure reaches dynamic equilibrium, no longer undergoing significant mergers \citep{Cuesta_etal}. Although this radius is challenging to define in realistic systems, owing to its proximity to galaxy cluster cores, it is generally associated with the zone where the intracluster medium is in hydrostatic equilibrium. The splashback radius, marking the point where infalling particles create a caustic at the first apocenter of their orbits, has recently gained attention in both simulations \citep{Splashback1, Splashback4, Splashback2, Splashback3, A_better_way_to_define_dark_matter_haloes} and observational studies \citep{Splashback_obs1, Splashback_obs2, Splashback_obs3, Splashback_obs4, Splashback_obs5} as an important measure for delineating cluster boundaries and assessing their accretion rate.

The turnaround radius is another critical scale in the study of halo formation, marking the transition from the collapsing structure to the Hubble flow. Within the spherical collapse model (SCM) framework \citep{GG1972}, turnaround represents the outermost boundary of halos occurring at a fixed density with respect to the background with its value being sensitive to the dark energy content of the Universe \citep{Turnaround_graveyard0, Turnaround_graveyard1}. 

Building on these insights, in our previous work \citep{KorkidisEtAl, Korkidis_etal2023} we showed that the density at the turnaround radius in realistic simulated clusters closely matches the predictions of the SCM. In \cite{PavlidouEtal2020}, we showed that a measurement of the present-day average matter density and its evolution with time at the turnaround scale for a few thousand clusters would be sensitive to the overall matter density in the universe and to the value of the cosmological constant. Such measurements have the potential to provide constraints on cosmological parameters that complement those obtained from Baryon Acoustic Oscillations (BAO), Type Ia Supernovae (SNIa), and the Cosmic Microwave Background (CMB) \citep{BAO1, SNIa, CMB-WMAP1}.

Despite its potential to introduce new constraints on the cosmological parameters, the study of turnaround in clusters remains under-explored. This is primarily due to the assumption that its observation requires access to the phase space information of galaxies surrounding clusters — a task complicated by redshift-space distortions and projection effects. Consequently, there is a prevailing notion that the turnaround scale is presently unobservable, which has, in turn, discouraged systematic searches for turnaround signals in the density profiles of cluster outskirts.

In contrast to the complexities involved in acquiring phase space information, density profiles are notably easier to observe in actual systems. On that front, the past decade has seen substantial progress in our ability to analyze the outer densities of galaxy clusters. Observations like the profile of satellite galaxies \citep{Satellite_obs1, Satellite_obs2}, Sunyaev-Zel’dovich effect and weak lensing \citep{WL_obs1, WL_obs2, WL_obs3, WL_obs4, WL_obs5} have now become commonplace for examining the outer regions of galaxy clusters. Inspired by these advancements, this work seeks to re-examine density profiles with a particular focus on the turnaround scale.

Our understanding holds, that the turnaround radius should reside beyond the splashback feature in density profile, provided there is a considerable infalling component surrounding the halos under consideration. At these scales, analytical models of density profiles, particularly those assuming spherical symmetry \citep{PradaEtal, TavioEtal, Betancort-RijoEtal06, Cuesta_etal}, have been very successful in describing stacked cluster halos in simulations and have demonstrated their universality up to the onset of significant shell-crossing. In particular, \cite{PradaEtal} showed that beyond approximately three Virial radii (which closely corresponds to the turnaround radius at low redshifts in simulations), the effect of shell crossing diminishes significantly. 

Motivated by these findings, in our recent work \citep{Korkidis_etal_2024} we achieved similar success in analytically modeling the outer density profiles of simulated clusters. In this study, we used excursion set theory for the statistical representation of initial profile distributions of halos, and the spherical collapse model (excluding shell-crossing) was employed to follow their non-linear evolution. This approach effectively modeled the outer regions of average cluster density profiles as a function of their collapsed mass. Through this model, we successfully derived scaling relations connecting the collapsed mass of clusters with their turnaround mass, achieving accuracy better than 10\% across various cosmologies.

These findings suggest that the SCM without shell-crossing may offer insights into identifying signatures of the turnaround radius within stacked cluster density profiles. Therefore, in this work, we aim to model the outer profiles of clusters and compare them with simulations to understand how deviations correlate with the turnaround radius. To accomplish this, we extend the methodology introduced in \cite{Korkidis_etal_2024} to develop an analytical description of the density profile as a function of its radial distance.


This paper is organized as follows: Section ~\ref{data} briefly outlines the simulation and halo sample used in this study and details the methodology for constructing density profiles. Section ~\ref{section: model} describes how the radial density profile is derived from the mass profile of \cite{Korkidis_etal_2024}. In Section ~\ref{section: results}, we compare the profiles and their logarithmic slopes of our analytical model against those of the simulated profiles, demonstrating that the Hubble flow leaves a recognizable mark on the profiles. Finally, section section ~\ref{conclusions} summarizes our findings.



\section{Simulated data} \label{data}


\subsection{Simulation overview and halo selection}

To analyze the stacked density profiles of cluster sized halos we used one of the dark-matter-only boxes from the second run of the MulitDark Planck simulations (MDPL2). On the scales of interest in this work baryonic effects have a negligible impact, thus a dark-matter-only box was sufficient for our needs. The simulated box that we employed used a Planck  $\Lambda$CDM cosmology \cite{Plank_cosmo_params} with a box of \(\rm 1000 \ cMpc/h \) on a side and particle mass of $\rm 1.51 \times 10^9 h^{-1} {\rm M_\odot}$. The halo catalog was produced with the Rockstar algorithm \citep{Rockstar}.

For the halo selection we considered the same two samples as in \cite{Korkidis_etal2023}: one sample with the 3000 most massive halos in the range \(\rm 10^{14} M_{\odot} \leq M_{200} \leq 6 \times 10^{15} M_{\odot}\) and another sample for which we randomly selected780 halos from logarithmically spaced mass bins, with \(\rm M_{200} \geq M_{min}\). We opted for a random halo selection for our second sample to keep the computational costs of the analysis tractable. The value of \(\rm M_{min}\) at  each redshift snapshot scaled proportionately to the mass of the largest halo at that time, and was equal to \(\rm 6 \times 10^{13} M_{\odot}\) at a \(\rm z=0\). The resulting median \(\rm M_{200}\) were 7.4, 3.8 and 1.8 \(\rm \times 10^{14} \ M{\odot} \) for z = 0, 0.5 and 1 respectively. Our choice to analyze the largest halos at each redshift, instead of making a selection based on the halos merger history, or accretion rate was mainly related to the fact that these parameters are not expected to affect the profiles at the scales of interest in this work. 

\subsection{Building the profiles}

To generate the profiles, we divided the area around each galaxy cluster's center into 500 concentric shells, extending to a radius of  \(\rm 25 Mpc\), ensuring that each halo was contained within the simulation's boundaries. The size of our bins were large enough to contain at least \(\rm 6 \times R_{200} \) in radius for both group and cluster-sized halos. Within the shells, we calculated the dark matter density and average radial velocities of dark matter particles. Following the methodology proposed by \cite{KorkidisEtAl}, we identified the kinetically driven turnaround scale as the radius \(\rm R_{ta}\) of the outermost shell that is not expanding, measured from the cluster's center. Then we made sure to discard halos that lied withing the turnaround radius of a bigger structure. Finally, we normalized the radius of each individual halo's density profile to its turnaround radius \(\rm R_{ta}\). 

\section{Analytic profile} \label{section: model}

In Section 2 of \cite{Korkidis_etal_2024} we employed excursion set theory for a statistical description of dark matter halos density and environment, and complimented it by the SCM to follow the non-linear growth of these structures. This process allowed us to derive a simple expression for the mode density profile of cluster sized halos as a function of their collapsed mass $\rm m$. Specifically, we established that the average matter density of a sphere enclosing mass $\beta m$ (so $\beta >1$ always) around a cluster-sized-halo, is universal, and equal to
\begin{equation} \label{enclosed density}
    \rho_{\rm avg} (\beta)=  \rho_m (1-\beta^{-\gamma})^{-\ed_c}\,,
\end{equation}
where $\rm \rho_m$ is the average matter-density of the Universe at the desired redshift; \(-\gamma = d\ln S /d \ln m \) is the logarithmic slope of the density field variance $S(m)$ when smoothed on a scale that encloses mass $m$; and $\ed_{\rm c}$ is the overdensity of a collapsing structure linearly extrapolated to its time of collapse. For detailed recipes to calculate \(\gamma\) and $\ed_{\rm c}$ refer to Appendices A and B of \cite{Korkidis_etal_2024}.

From Eq. ~\ref{enclosed density}, it is straightforward to derive the local density within a narrow spherical shell of radius\footnote{Here, \(\rm R_{ta}\) is the turnaround radius, but it could have been the radius of a sphere enclosing an arbitrary multiple of the average background-matter density of the Universe as long as it is lower than the virial (collapsed-structure) overdensity, usually taken to be equal to 200. For a detailed discussion on the matter see \cite{Korkidis_etal_2024}.} \(\rm \widetilde{R} \equiv R/R_{ta} \).  A detailed derivation can be found in Appendix ~\ref{appendix A}.  
The functional form of this radial density profile is given parametrically by the following two equations:
\begin{eqnarray} 
    \rho(\beta) &=& \rho_{m}\left[1-\beta^{-\gamma}\right]^{-\ed_c+1}\,,
\label{density shell} \\
    \rm \widetilde{R^3}(\beta) &=& (1+\delta_{ta})
    \left[
    1 - \left(1+\delta_{ta}\right)^{-1/\ed_c}
    \right]^{1/\gamma} 
    \beta \left[1- \beta^{-\gamma}\right]^{\ed_c}\,.
\label{radius mass}
\end{eqnarray}
Equation (\ref{density shell}) gives the local density on the surface of a sphere enclosing mass \(\beta m\) and Eq. ~\ref{radius mass} connects the mass enclosed by the sphere with its normalized radius. Since Eq. ~\ref{radius mass} cannot be solved for $\beta$ in closed form,  it has to be solved numerically or be approximated by a simpler function.

Importantly, equations ~\ref{density shell} and  ~\ref{radius mass} describe the local density profile as a function of the turnaround radius and the cosmological parameters on which \(\ed_c , \ \gamma \ \rm{and} \ \rm{\delta_{ta}} \ \) depend. From \cite{PavlidouEtal2020} and \cite{Korkidis_etal2023} we know that the turnaround density and its evolution with cosmic time are sensitive to the value of the cosmological constant. This is true both in the context of spherical collapse and in simulations. In our formalism, the dependence on redshift when the profile is expressed in terms of $R_{\rm ta}$ enters through $\delta_{\rm ta}$.  This introduces a dependency of the profiles not only on the present-day matter density (\(\rm \Omega_m\)) but also on the value of the dark energy density \(\rm \Omega_{\Lambda}\).

\section{Comparison between simulated and analytic profiles} \label{section: results}


It is widely acknowledged that density profiles exhibit universal characteristics when expressed in units of a 
constant-overdensity radius. Thus, since the kinematically defined turnaround radius corresponds to a constant overdensity as described by the SCM \citep{KorkidisEtAl, Korkidis_etal2023}, it is reasonable to anticipate that our simulated profiles, when normalized with respect to this radius, will exhibit similar behavior.

\begin{figure}[htb!] 
    \includegraphics[width=1\columnwidth]{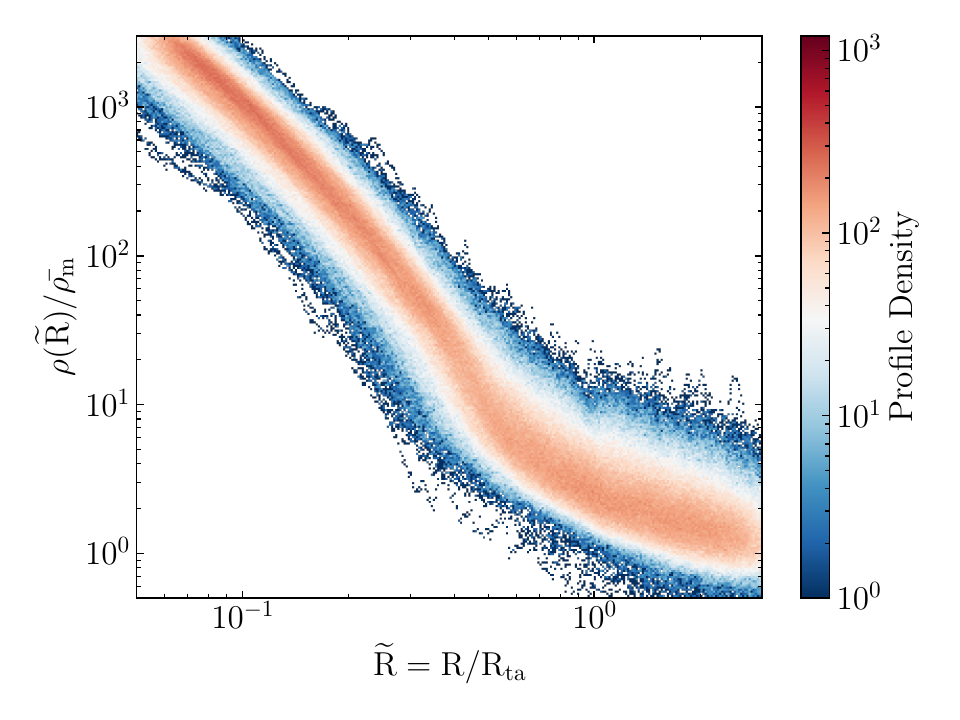}
    \caption{Dark matter density (in units of the background-Universe average matter density) plotted against the normalized radius of concentric spherical shells around MDPL2, $z=0$ halos. The color scale indicates the number of lines per pixel on the plot. The radii of the shells has been normalised with respect to the kinetically defined turnaround radius.}
    \label{figure: bundle}
\end{figure}

In Fig.~\ref{figure: bundle} we plot the normalized density as a function of the normalized radius of our halo sample at present time. The density and radii have been normalized with respect to the mean matter density of the universe and the kinematically-defined turnaround radius, respectively. From the density of lines (represented by color in the plot) it is evident that the majority of individual profiles aggregate in a narrow band. Around \(\sim 0.4R_{ta}\) there is a steep break in the slope of this band. Further out, the slope increases smoothly (from negative to zero) as the profiles will eventually asymptote to the mean matter density of the Universe. It is this behavior - the gradual change in slope - that we seek to capture with the analytic description of the outskirts of clusters discussed in the previous section.

To make meaningful comparisons between simulations and our analytical predictions, we grouped these profiles into 30 linear radial bins, from 0 to \(3\times R_{ta}\), and within each bin we calculated the median value of the density, thus constructing the "median profile". We then computed the logarithmic slope of this median profile using finite differences and smoothing them with a Savitzky–Golay filter with a 5-point window and a 3rd order polynomial\footnote{Since the Savitzky–Golay filter fits a polynomial, it can also be used for direct derivative calculation. We have verified that this approach yields results similar to first computing the derivative and then applying smoothing.}. 

\begin{figure}[htb!]
\includegraphics[width=1\columnwidth]
{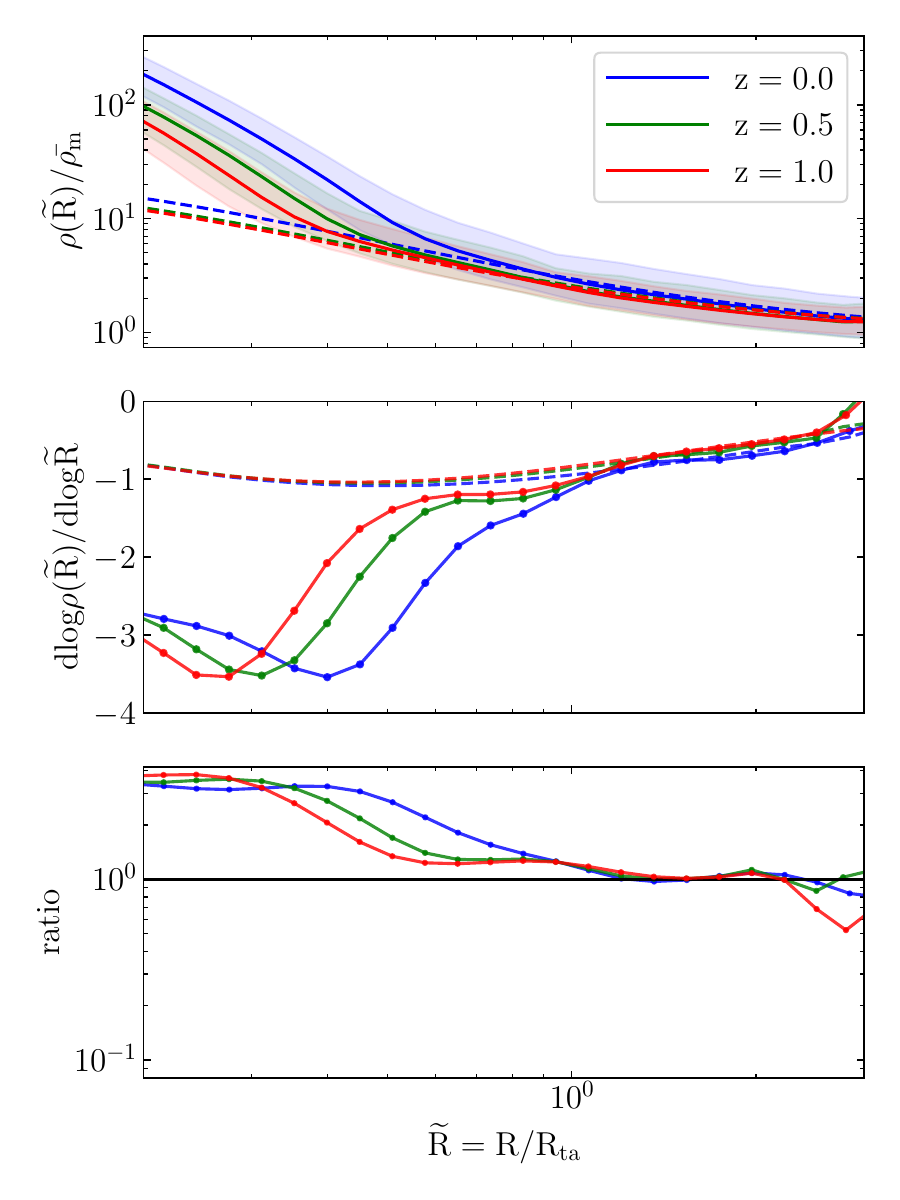}
    \caption{Comparison between simulated and analytic density profiles for different redshifts. Upper panel: solid lines depict the median density profile at different redshifts of our MDPL2 sample in spherical shells of radius \(\rm R/R_{ta}\); shaded regions represent the \(\rm 1\sigma\) spread of the density within each shell; dashed lines represent the analytical prediction for the outer portions of the mode profile discussed in section ~\ref{section: model}. Middle panel: Logarithmic slope of the density profiles, with dashed lines illustrating the logarithmic slope of the analytical prediction. Lower panel: ratio of logarithmic slope in median simulated profile over the one of the analytically predicted profile. The analytic slope deviates from the one in simulations for $R/R_{\rm ta} \lesssim 1$.
    \label{figure: simualted VS analytic}}
\end{figure}

In the upper panel of Fig.~\ref{figure: simualted VS analytic}, we overlay our analytical model with the median density profiles at various redshifts of the simulation. In the middle panel we display their logarithmic slope. Since our primary focus is the profile slopes, we opted to compare our analytic profile (which models the mode density) with the median because the different representations mainly affect the profile normalizations. Additionally, calculating the mode would necessitate smoothing the profiles, which can introduce difficult-to-track correlations with the various cluster masses in each radial bin. Finally, the median has the benefit of being more readily calculable and having easily defined errors. For a comparison between different profile statistics, we refer the reader to Appendix ~\ref{appendix B}.

From the figure, three main observations emerge:

(A) Starting at the turnaround radius and for larger radii (\(\rm R/R_{ta} \gtrsim 1\)), both the profile shapes and their normalization remain largely unchanged, with the slope increasing smoothly as it reaches for the constant background density. From \(\rm z=0 \ to \ z=1\), we anticipate significant variations in collapsed masses, accretion rates, and concentrations, all of which impact the central cluster regions differently. However, despite these changes in the dynamical state of our halo sample, the outer profiles remains universal and the turnaround is  reached close to a slope value of -1.

(B) The shapes of the median profiles at large scales, where the Hubble flow dominates, are well described by the physics encoded in early Universe statistics and spherical collapse without the assumption of shell crossing. This is evident from the excellent capture of profile slopes by our analytical model, as depicted in the middle and lower panels of the figure. In contrast, these assumptions fail profoundly within the infall region (\(\rm R/R_{ta} < 1\)) of the clusters. In these regions, as we approach the cluster cores, the profiles are shaped by a complex interplay of different particle trajectories surrounding the central cluster regions. As expected, this complexity surpasses what the simple assumption of simple spherical collapse can account for. A more involved approach to model this situation would at least require introducing the concept of shell crossing (see for example \cite{PradaEtal06}).

(C) Redshift predominantly influences the inner regions of the density profiles \(\rm R/R_{ta} \leq 0.4\). This is evident in the alteration of the profile normalization at small radii, as well as in the position of their splashback feature. The latter is represented by the steepening of the slope profile seen in the middle panel. Despite our simple criterion for sampling the largest structures in the simulated box, the variation in profile normalization can not be attributed to the decrease in the average \(\rm M_{200}\) of our halo sample. Rather, this change is mostly driven by the fact that the average \(\rm R_{ta}\) of clusters decreases with redshift and is associated with a higher overdensity with respect to the background Universe. 

The divergence between analytical and simulated profiles near the turnaround radius, coupled with the universality of profiles beyond this radius, has a significant implication for the observability of the turnaround radius: the turnaround scale is imprinted in 3-dimensional profiles as the scale beyond which the average/median/mean profile slope\footnote{The statement is true for all three statistics; see Appendix \ref{appendix B}} matches the one expected from early-Universe statistics and spherical collapse. 
We expect this feature to continue to be true in observable projected (column density) profiles. 
Projected density profiles observed in the plane of the sky are contaminated at all scales by material unrelated to orbiting parts of clusters. Beyond the turnaround radius, however, even projected material should be well-described by our analytic profile.
Consequently, deviations between analytical modeling of the outer profiles and actual projected density profiles, similar to this work, could be closely connected to the location of the turnaround radius.

\begin{figure*}[htb!] 
    \includegraphics[width=1\textwidth]{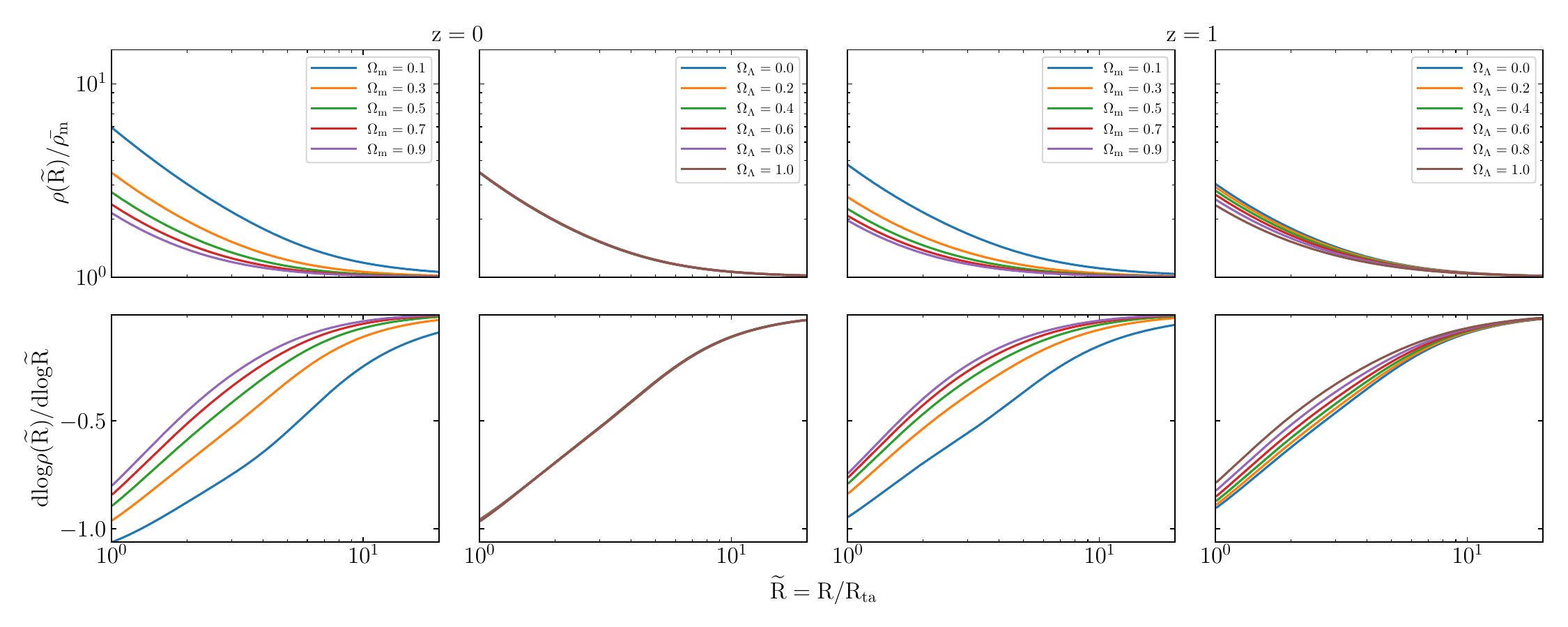}
    \caption{Dependence of the analytical profile and its logarithmic slope on the cosmological parameters. Upper panels illustrate matter density across shells, normalized by the turnaround radius, at redshifts indicated by the titles. With different line colors we show how the profiles change for different values of \(\rm \Omega_m\) (assuming \(\Omega_{\Lambda}=0.7\)) and \(\Omega_{\Lambda}\) (assuming \(\Omega_{m}=0.3\)). Bottom panels explore the change in the logarithmic slope of the respective density profiles.}
    \label{figure: profile dependencies}
\end{figure*} 

Another implication is related to the significant impact of the turnaround radius redshift evolution on the variation of profile normalizations. As this evolution is influenced by the values of the cosmological parameters \citep{PavlidouEtal2020, Korkidis_etal2023}, we can expect that this dependence will also be reflected in the profiles. In Fig.~\ref{figure: profile dependencies} we investigate this hypothesis by plotting the analytic profile and its logarithmic slope for different values of the cosmological parameters. To construct the profiles we assumed a Hubble constant \(\rm h_0=0.7\), a power spectrum slope \(\rm n_s=0.96\) and median virial mass for the cluster of \(\rm M_{200}=10^{14} M_{\odot}\). Then, we used the scaling relation of Eq. 11, introduced in \cite{Korkidis_etal_2024} and the known connection between overdensity masses and radii to derive the SCM turnaround radius for clusters with the quoted virial mass (\(\mathbf{\rm M_{200}}\)). The small error associated with this relation makes it a good proxy for our lack of access to the true turnaround radius of the individual clusters that would make these profiles (see Fig. \ref{figure: Rta scaling with R200}, Appendix ~\ref{appendix D}).

Looking at the first two columns of the figure, depicting the model at \(\rm z=0\), we observe a pronounced dependence on the Universe's overall matter content. This effect is particularly prominent at low redshifts but slightly decreases as we approach \(\rm z=1\).  While the profile exhibits no discernible change with varying values of the cosmological constant at \(\rm z=0\), this effect becomes more pronounced with increasing redshift (third and fourth columns of the figure). Hence, the change in the profile sensitivity to \(\rm \Omega_{\Lambda}\) with redshift should be attributed to the turnaround radius. This is because, as the cosmological parameters and redshift change, so does \(R_{ta}\) as it corresponds to a different overdensity as predicted by the SCM. Had we kept the overdensity at a fixed value for all redshifts and cosmologies, we would have mitigated this effect. This is clearly demonstrated in Fig. ~\ref{figure: profile dependencies R200} of Appendix ~\ref{appendix C}, where we illustrate the dependence of the profiles on the cosmological parameters when normalizing the analytic profile with respect to \(\rm R_{200}\). 

This observation opens a significant avenue to impose constraints on the cosmological parameters through the turnaround density without directly observing the turnaround radius of individual clusters. Just as we used the scaling relation \ref{scaling relation} to find the turnaround radius from \(\rm R_{200}\), which were then used to normalize the profiles of Fig.~\ref{figure: profile dependencies}, we could similarly apply these scaling relations on observed stacked clusters at different redshifts. Fitting our model to the outer portions of these profiles could, in principle, allow us to derive constraints on the cosmological parameters, while the scale where their slopes start to deviate from universality is expected to correlate strongly with the average turnaround radius of the clusters.

This is of course a statement of principle. Applying this technique to data requires a careful assessment of the systematics involved in simulated surveys and mock observations.

\section{Conclusions} \label{conclusions}

Despite the extensive research on cluster density profiles, a gap remains in our modeling of their transition to the Hubble flow. To address this gap, we normalized stacked cluster-density profiles from simulations to their turnaround radius which represents the scale at which radial velocities around clusters, on average, become zero. Subsequently, we compared these profiles with an analytical model for the most probable outer radial density profile. This model, a continuation of the work by \cite{Korkidis_etal_2024}, is based on Gaussian early Universe statistics and the spherical collapse model without shell crossing and is found to be sensitive both to the overall matter content of the Universe and the cosmological constant \((\rm{\Omega_m \ and \ \Omega_{\Lambda}} \)).

From our analysis, we find the simulated profiles to be universal in shape and well-described by our analytical model beyond turnaround, where the Hubble flow dominates. However, notable deviations are observed within the turnaround radius, signaling the emergence of shell-crossing effects. At smaller scales, after a sudden steepening of the profile slopes at the splashback radius, this universality breaks down. The profiles exhibit variations correlated with changes in their dynamical state, as demonstrated by profile changes at various redshifts within our halo sample. These variations impact both the shapes and normalization of the profiles, challenging the validity of the assumptions made by our model.

Our findings suggest that analytical, spherically symmetric models of halo formation without shell-crossing could be instrumental in identifying the turnaround radius of stacked clusters in observational data. Deviations from such models in stacked density profiles are anticipated to closely correlate with the position of their average turnaround radius. Furthermore, the ability of these models to accurately describe the outer regions of density profiles is expected to be sensitive to the values of cosmological parameters.

\begin{acknowledgements}

We acknowledge support by the Hellenic Foundation for Research and Innovation under the “First Call for H.F.R.I. Research Projects to support Faculty members and Researchers and the procurement of high-cost research equipment grant”, Project 1552 CIRCE (GK, VP); by the European Research Council under the European Union's Horizon 2020 research and innovation programme, grant agreement No. 771282 (GK); and by the Foundation of Research and Technology – Hellas Synergy Grants Program (project MagMASim, VP). 
The research leading to these
results has received funding from the European Union’s Horizon 2020 research and innovation programme under the Marie Skłodowska-Curie RISE action, Grant
Agreement n. 873089 (ASTROSTAT-II).
The CosmoSim database used in this paper is a service by the Leibniz-Institute for Astrophysics Potsdam (AIP). The MultiDark database was developed in cooperation with the Spanish MultiDark Consolider Project CSD2009-00064. The authors gratefully acknowledge the Gauss Centre for Supercomputing e.V. (www.gauss-centre.eu) and the Partnership for Advanced Supercomputing in Europe (PRACE, www.prace-ri.eu) for funding the MultiDark simulation project by providing computing time on the GCS Supercomputer SuperMUC at Leibniz Supercomputing Centre (LRZ, www.lrz.de). The Bolshoi simulations have been performed within the Bolshoi project of the University of California High-Performance AstroComputing Center (UC-HiPACC) and were run at the NASA Ames Research Center. Throughout this work we relied extensively on the PYTHON packages Numpy \citep{Numpy}, Scipy \citep{SciPy} and Matplotlib \citep{Matplotlib}.

\end{acknowledgements}

\bibliographystyle{aa}
\bibliography{bibliography}

\begin{appendix} 

\section{Derivation of the analytical outer density profile} \label{appendix A}

In this Appendix we use the profile for the average density $\rho_{\rm avg}$ in a sphere around a collapsed structure of mass $m$, enclosing mass $\beta m$, to derive the local density within a thin spherical shell. A shell of mass $md \beta$ and of volume $\rm 4 \pi R^{2}dR$ (where an increment of radius $dR$ corresponds to an increment in mass $md \beta$) will have local density given by the ratio:
   \begin{equation}\label{simple_ratio}
   \rho = md\beta/4\pi R^2 dR,
\end{equation}
whereas the average density within the entire sphere of mass $\beta m$ will be connected to this mass and radius through
   \begin{eqnarray} \label{beta and R}
\rho_{\rm avg} =3\beta m/4\pi R^3\,.
\end{eqnarray}
Differentiating this relation with respect to $\beta$ and substituting in Eq. ~\ref{simple_ratio} we obtain 
\begin{equation}\label{step1}
   \rho(\beta) = \frac{m}{4\pi R^2} \frac{3\beta}{R \left(1 - \frac{\beta}{\rho_{\rm avg}}\frac{d\rho_{\rm avg}}{d\beta}\right) } = 
   \frac{\rho_{\rm avg}}{1 - \frac{\beta}{\rho_{\rm avg}}\frac{d\rho_{\rm avg}}{d\beta}}\,.
\end{equation}
To calculate $(\beta/ \rho_{\rm avg})(d\rho_{\rm avg}/d\beta)$, we differentiate Eq.~(\ref{enclosed density}) with respect to \(\beta\) to obtain: 
\begin{equation}
\frac{\beta}{\rho_{\rm avg}} \frac{d\rho_{\rm avg}}{d\beta} =
\frac{-\ed_c \gamma \beta^{-\gamma}}
{\left[1-\beta^{-\gamma}\right]}\,.
\end{equation}
Substituting this equation in Eq.~(\ref{step1}) we obtain
\begin{equation}\label{miracle}
      \rho(\beta) = \frac{\rho_{m,0}a^{-3}\left[1-\beta^{-\gamma}\right]^{-\ed_c+1}}
      {1+ (\ed_c\gamma -1)\beta^{-\gamma} }\,.
  \end{equation}
The product $\ed_c\gamma$ for a concordance \(\Lambda \rm CDM\) cosmology and for cluster masses \(M_{200}\geq 6 \times 10^{13} M_{\odot}\) has a value very close to 1.
Thus, for \(\beta>1\), which concerns us in this work, the entire second term in the denominator will be suppressed, and we can write the much simpler expression 
\begin{equation}\label{simple}
    \rho(\beta) = \rho_{m,0}a^{-3} \left[1-\beta^{-\gamma}\right]^{-\ed_c+1}\,.
\end{equation}
This relation connects the density in a thin shell at the surface of a sphere with its enclosed mass \(\beta m\).
We next turn to expressing this profile in terms of $R/R_{\rm ta}$. We normalize Eq.~(\ref{beta and R})
 to the turnaround scale and use again Eq.~(\ref{enclosed density}) for $\rho_{\rm avg}$ to obtain
\begin{equation} \label{R and beta}
    \left(\frac{R}{R_{ta}}\right)^3 = 
    \frac{\beta  \left[1- \beta^{-\gamma}\right]^{\ed_c} \rho_{\rm ta}}{\beta_{\rm ta}\rho_{m,0}a^{-3}} = 
\frac{\beta}{\beta_{\rm ta}}  \left[1- \beta^{-\gamma}\right]^{\ed_c}(1+\delta_{\rm ta}) ,
\end{equation}
where $\beta_{ta}$ is the turnaround mass of a structure in units of its collapsed mass,  $\rm \rho_{ta}$ is the turnaround density predicted by the SCM, and $1+\delta_{\rm ta} = \rho_{\rm ta}/\rho_{m}(a)$ is the density contrast of a structure at turnaround. For $\beta_{ta}$, Eq.~(\ref{enclosed density}) gives
\begin{equation} \label{beta_ta}
    \beta_{ta} = 
    \left[
    1 - \left(1+\delta_{\rm ta}\right)^{-1/\ed_c}
    \right]^{-1/\gamma}\,.
\end{equation} 
Substituting Eq.~\ref{beta_ta} in Eq. ~\ref{R and beta} we obtain
\begin{equation}\label{irreversible}
    \widetilde{R}^{\,3} = (1+\delta_{ta})
    \left[
    1 - \left(1+\delta_{ta}\right)^{-1/\ed_c}
    \right]^{1/\gamma} 
    \beta \left[1- \beta^{-\gamma}\right]^{\ed_c} 
\end{equation}
Equation ~\ref{R and beta} cannot be solved for $\beta$ in closed form. Thus our options are to either find $\beta=\beta(\widetilde{R})$ numerically or approximate it by a polynomial. An example of such an approximation would be to take a Taylor's expansion around $\widetilde{R}=1$ (the turnaround radius) at a desired accuracy; we find that beyond second order of the expansion mostly affect the profile at radii smaller than \(0.3R_{ta}\) where the profile fails by design. Throughout this work we take the latter approach.

\section{The radial density profile through different statistics} \label{appendix B}
\begin{figure}[htb!] 
    \includegraphics[width=1\columnwidth]{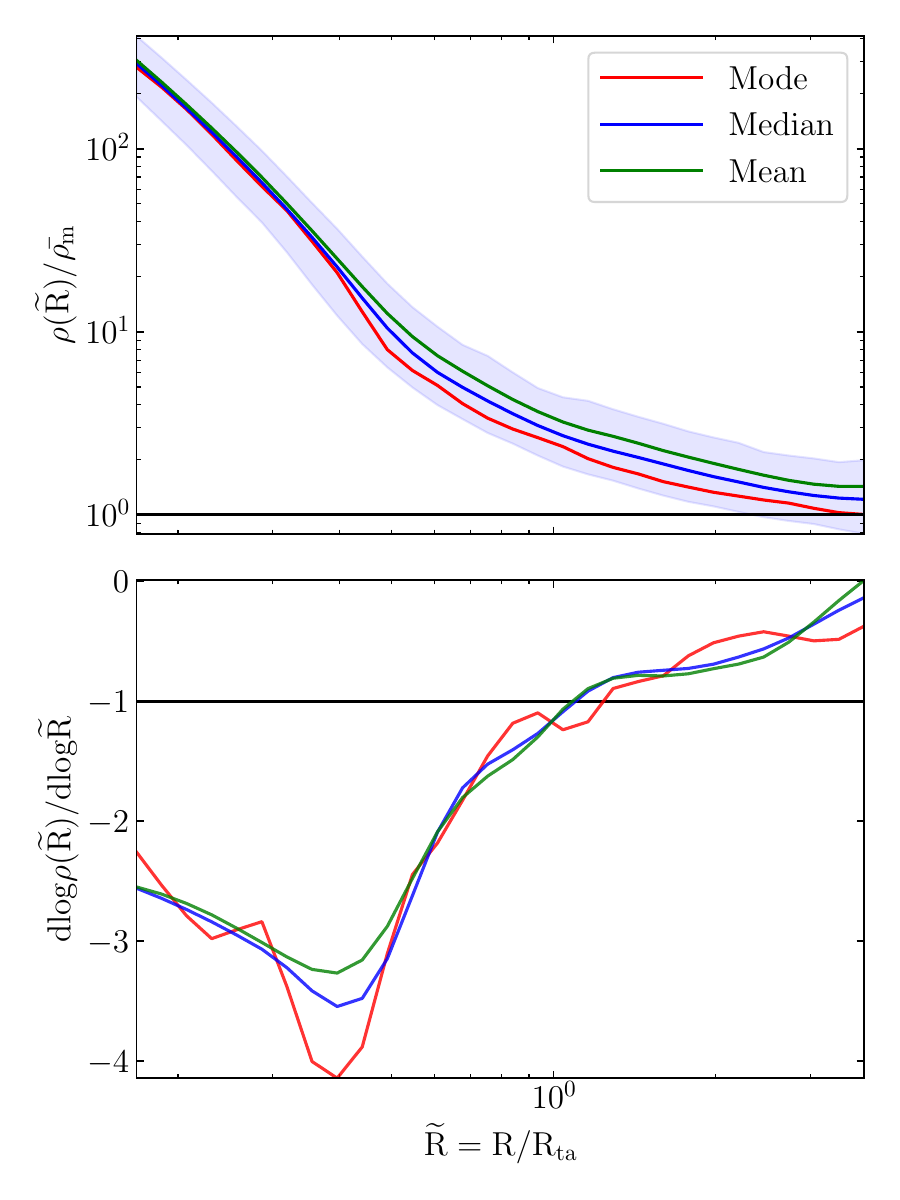}
    \caption{The effect of choosing different statistics to represent the density profile. The two panels present the mode, median and mean density profiles along with their logarithmic slopes of our halo sample at \(\rm z=0\).}
    \label{figure: simualted different stats}
\end{figure}

In the upper panel of figure ~\ref{figure: simualted different stats}, we plot the mode, the median, and the mean profiles (local density in radial shells, normalized to the mean matter density of the Universe, as a function of  shell radius, normalized to the turnaround radius) for our $z=0$ sample. 
To calculate the different the profiles from different statistics we followed the procedure described in section ~\ref{section: results} for the median. That is, we grouped the profiles into 30 linear radial bins, from 0 to \(\rm 3\times R_{ta}\) and within each bin we calculated the mode, median and mean value of the density. For the mode we employed a Gaussian kernel density estimator \citep{SciPy} to model the probability density function (PDF) of densities in the bin. The mode corresponded to the density for which the PDF has a maximum.
The shaded region corresponds to the $1\sigma$ spread of profiles. In the lower panel we plot the logarithmic slope of each profile. Further out of the splashback feature, different profiles exhibit different normalization (highest for the mean, lowest for the mode). However, the slopes of the profile for radii beyond the turnaround radius appear mostly unaffected by the type of statistic used to form the profile. As expected, the mode profile is more noisy than the median or the mean.

\section{Scaling relation between $
\rm R_{ta}$ and $\rm R_{200}$} \label{appendix D}

In \cite{Korkidis_etal_2024} we demonstrated that our model for the most probable outer density profile of galaxy clusters allows for the derivation of analytical mass scaling relations that depend on cosmology and redshift. These relations connect cluster masses at different overdensities, including the turnaround, and were shown to perform well, with the peak of the error distribution remaining below 15\% across all cosmologies, mass samples, and overdensities tested.

Since the kinetically defined turnaround radius corresponds to a constant overdensity as described by the SCM, we can use the known connection between overdensity masses and radii (corresponding to overdensities \(\rm \delta=200 \ and \ \delta=\delta _{ta},\) respectively) along with the scaling relation of to show 

   \begin{equation}\label{scaling relation}
   \rm R_{ta} = \left ( \rho_{200} \beta_{ta} /\rho_{ta}  \right )^{1/3} \times R_{200}.
    \end{equation}
    Here $\rm \beta_{ta}$ is the scaling factor between the "collapsed mass" $\rm M_{200} \ and \ M_{ta}$ (Eq. 11, \cite{Korkidis_etal_2024} for $X \equiv X_{ta}$) while \(\rm \rho_{200}\) is the density at \(\rm \delta=200\).

\begin{figure}[htb!]
    \includegraphics[width=1\columnwidth]{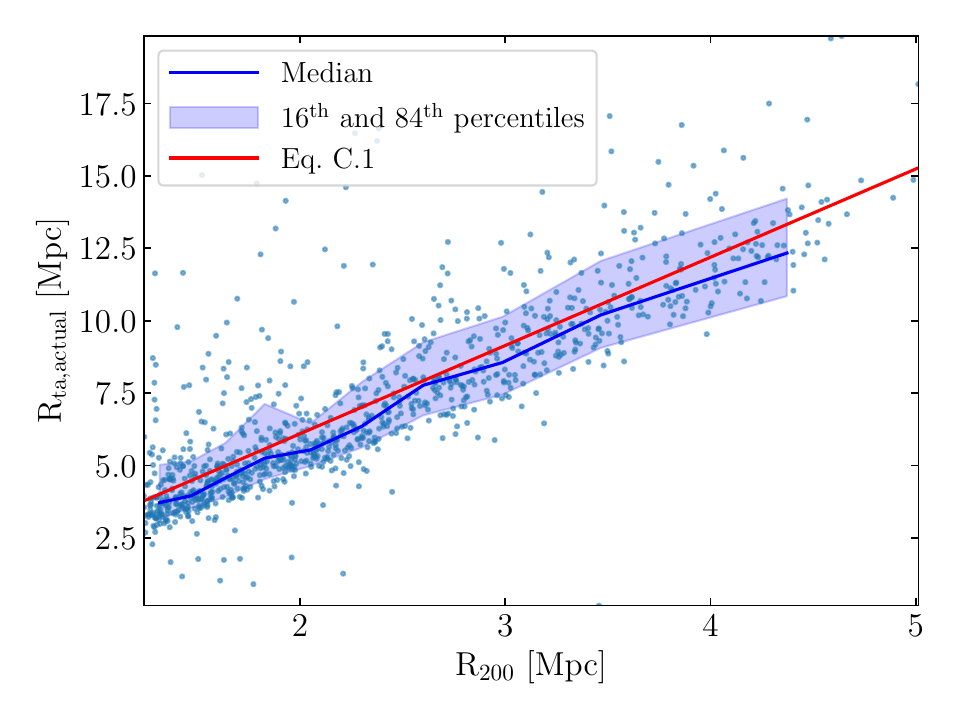}
    \caption{Correlation between the true turnaround radius \(\rm R_{ta,actual}\) and \(\rm R_{200}\) for our cluster sample at \(\rm z=0\). The red solid line depicts the theoretical scaling relation of Eq. ~\ref{scaling relation}. The blue solid line represents the median value of the blue points in bins of \(\rm R_{200}\). The blue shaded region corresponds to the 16th and 84th percentiles.}
    \label{figure: Rta scaling with R200}
\end{figure}

In Fig. ~\ref{figure: Rta scaling with R200}, we test that this scaling relation exists in simulations by plotting the actual turnaround radius \(\rm R_{ta,actual}\) as a function of \(\rm R_{200}\) for our MDPL2 cluster sample at \(\rm z=0\). The median along with the 16th and 84th percentiles of this scaling in bins of \(\rm R_{200}\), is shown with the solid blue line and the shaded region, respectively. When comparing the median with the scaling relation in Eq. ~\ref{scaling relation}, depicted in red,we observe a strong correlation between the two, with a median offset close to 5\%.

\section{Cosmological dependence of outer profile when normalized to $R_{200}$} \label{appendix C}

\begin{figure*}
    \includegraphics[width=1\textwidth]{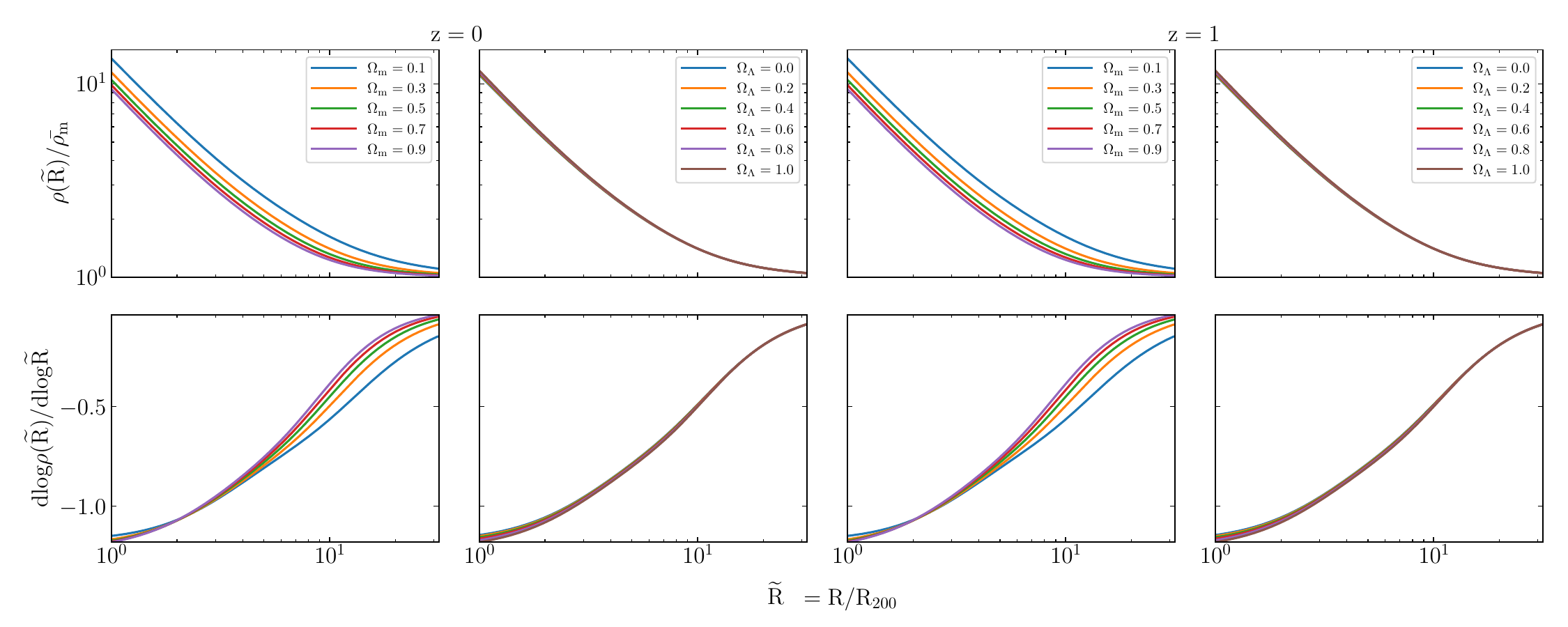}
    \caption{Analytical profile dependencies on the cosmological parameters at different redshifts. As in Fig. ~\ref{figure: profile dependencies}, different panels show the change of the density profile and its logarithmic slope with the difference that in this figure the radii have been normalized with respect to \(\rm R_{200}\).}
    \label{figure: profile dependencies R200}
\end{figure*} 

Figure \ref{figure: profile dependencies} illustrates that the most probable outer density profile of Section \ref{section: model}, when normalized with respect to \(\rm R_{ta}\) exhibits a dependence on \( \rm \Omega_{\Lambda}\) that evolves with redshift. However, of the three model parameters \(\ed_c , \ \gamma \ \rm{and} \ \rm{R_{ta}} \ \)only the turnaround radius has a redshift dependency, as it corresponds to the overdensity at which the SCM predicts turnaround to occur. It is thus reasonable to expect this sensitivity to \(\rm \Omega_{\Lambda}\) to stem from \(\rm{R_{ta}}\).

To test this hypothesis, in \ref{figure: profile dependencies R200} we plot our model and its logarithmic slope as a function of \(R_{200}\), for different values of the cosmological parameters and redshift. From the second column of this figure, it is evident that by keeping the radius at a constant overdensity (is this case, 200), we eliminate any dependence the profile would otherwise have on \(\rm \Lambda\). Simultaneously, we observe that the dependence on \( \rm \Omega_{m}\) does not evolve with redshift, in contrast to Fig. ~\ref{figure: profile dependencies}, where the sensitivity to \(\rm \Lambda\) come at the expense of the sensitivity to \( \rm \Omega_{m}\).



\end{appendix}

\end{document}